\documentstyle[prb,aps,psfig,epsfig,floats]{revtex}
\begin{document}
\twocolumn[\hsize\textwidth\columnwidth\hsize\csname@twocolumnfalse\endcsname
\title{Theoretical analysis of continuously 
driven dissipative solid-state qubits} 
\author{M.C.\ Goorden$^{1,2}$ and F.K. Wilhelm$^{1,3}$}

\address{$^1$ Quantum Transport Group,
Department of Applied Physics and DIMES, Delft University of Technology, 
Lorentzweg 1, 2628 CJ Delft, The Netherlands\\
$^2$ Instituut Lorentz, University of Leiden, P.O. Box 9506, NL-2300 RA
Leiden, The Netherlands\\
$^3$Sektion Physik and CeNS, Ludwig-Maximilians-Universit\"at,
Theresienstr. 37, 80333 M\"unchen, Germany}

\maketitle
\begin{abstract}
We study a realistic model for driven qubits using the
numerical
solution of the Bloch-Redfield equation as well as analytical approximations
using a high-frequency scheme. 
Unlike in idealized rotating-wave models suitable for NMR or quantum
optics, we study a driving term which neither is orthogonal to the static
term nor leaves the adiabatic energy value constant. We investigate the
underlying dynamics and analyze the spectroscopy peaks obtained in 
recent experiments. We show, that unlike in the rotating-wave case, this
system exhibits nonlinear driving effects.
We study the width of spectroscopy 
peaks and 
show, how a full analysis of the parameters of the
system can be performed by comparing the first and second resonance. We
outline the limitations of the NMR linewidth formula at low temperature
and show, that spectrocopic peaks experience a strong shift which goes much
beyond the Bloch-Siegert shift of the Eigenfrequency.
\end{abstract}
]

Coherent manipulation of quantum states is a well established technique
in atomic and molecular physics. In these fields, one works with
``clean'' 
generic quantum systems which can be very well decoupled from their 
environments. Moreover, it is possible to apply external fields in a way
such that strong symmetry relations between the static and the time-dependent
part of the Hamiltonian apply and the resulting dynamics is very simple
and can be treated analytically. In solid-state systems, the situation is
different. Not only do they contain a macroscopic number of
degrees of freedom which form a heat bath decohering the quantum states
to be controlled, but also is the choice of controllable parameters much
more restricted. A quantum-mechanical 
two state system (TSS) realized in a mesoscopic circuit can be identified with 
a (pseudo)spin, however, in that case the different components of the spin
may correspond to physically distinct observables such as e.g.\ magnetic
flux and electric charge \cite{Vion}. 
This naturally limits the possibilities of 
controlling arbitrary parameters of the pseudospin. Hence, in order to 
describe the direct control of quantum states in mesoscopic devices, 
concepts from NMR or quantum optics cannot be {\em directly} applied 
but have to be carefully adapted. In particular, as decoherence is usually 
rather strong in condensed matter systems, one can attempt to drive the
system rather strongly in order to have the operation time for a quantum gate,
usually set by the Rabi frequency, as short as possible.

We concentrate on the case of a persistent current quantum
bit\cite{Hans,Caspar,Irinel} 
driven through the magnetic flux through the loop and 
damped predominantly by flux noise \cite{CasparFrankEPJ} with Gaussian 
statistics. This setup is accurately described by the 
driven\cite{GrifoniHanggi,LHartmann} spin-boson model\cite{Leggett}
\begin{equation}
H=\frac{\epsilon(t)}{2}\hat{\sigma}_z-\frac{\Delta}{2}\hat{\sigma}_x
+\hat{\sigma}_z\sum_i c_ix_i + \sum_i \left(\frac{\hat{p}_i^2}{2m_i}
+\frac{1}{2}m_i\omega_i^2x_i^2\right) 
\label{eq:Hamiltonian}
\end{equation}
where $\epsilon(t)=\epsilon_0+s\cos\Omega t$
and the oscillator 
bath is assumed to be ohmic with a spectral density
$J(\omega)=\frac{\pi}{2}\sum_i \frac{c_i^2}{m_i\omega_i}\delta(\omega-\omega_i) =
\alpha\omega e^{-\omega/\omega_c}$. The connection of $J(\omega)$ to the
setup parameters is detailed in \cite{CasparFrankEPJ}. 
The static energy 
splitting of the peudospin is $\nu=\sqrt{\epsilon_0^2+\Delta^2}$. 
This model is  also 
applicable to other Josephson qubits and other
realizations \cite{Leggett,Weiss}. In particular,
the strong driving regime we are going to elaborate on has recently been
realized in several setups \cite{Alex,Andreas,Yasu}.
 We study the effective 
dynamics of the pseudospin having traced out the bath in 
the limit of weak damping, $\alpha\ll1$, which is appropriate
for quantum computation. This is done using the
Bloch-Redfield equation\cite{ArgyresKelley}.
The resulting equation is of Markovian form in the sense that it only
contains the density matrix at a single time, however, it
is derived 
in such a way, that the free coherent evolution
during the interaction with the bath 
is fully taken into account such that 
the resulting equation is numerically equivalent to a fully non-Markovian
path-integral scheme \cite{LHartmann,Weiss} and only memory terms
beyond the Born approximation are dropped. The
explicit form of the equations for this situation as well as the formulas
for the rates correspond to those given in \cite{LHartmann}. We compare our
numerical results to analytical formulae derived in the framework of
a high-frequency approximation\cite{Sasetti} which involves averaging
over the driving field and has nonetheless shown to give a good estimate
for the system dynamics even 
close to resonances\cite{LHartmann}. 

Initial experiments on quantum bits such as Ref. \onlinecite{Caspar} do not 
monitor the real-time dynamics of the system as in Ref.\ \onlinecite{Irinel}, 
because the read-out
is much slower than the decoherence, i.e.\ the dephasing time $\tau_\phi$
is too short. In order to optimize the experimental setup, it is 
important to measure both $\tau_\phi$ and the relaxation time $\tau_R$, even
and in particular if they are insufficient. In the standard NMR-case, this
is done by studying the width of the resonance \cite{Abragam}. We will 
detail, a somewhat modified analysis can be performed for solid-state qubits
and what are its limitations. We discuss both situations\cite{Caspar,Irinel}. 
Our results thus help to analyze the
decoherence as observed in Refs.\ \onlinecite{Caspar,Irinel}, and outline
the possibilities and limitations of driving the system in the nonlinear
regime. 
\begin{figure}[htb]
\includegraphics[width=0.9\columnwidth]{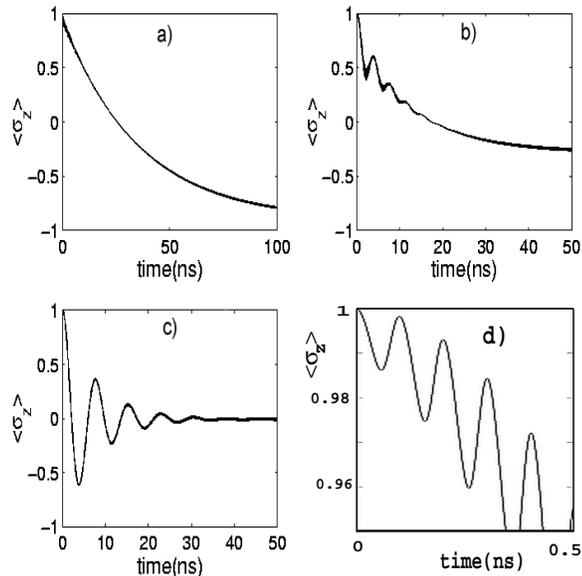}
\caption{\label{fig:rabiflop}
$\langle\sigma_z\rangle (t)$ at fixed frequency $\Omega/2\pi=6.6$GHz
 for different
bias points. a) Off-resonance, $\nu/2\pi=2.9$GHz, incoherent decay towards thermal
equilibrium; c) on-resonance, $\nu/2\pi=6.6$GHz, Rabi oscillations decaying
on the scale of a dephasing time $\tau_\phi$ towards a dynamical equilibrium
state; b) close to the resonance, $\nu/2\pi=6.4$ GHz, 
combination of decoherence
and relaxation; d) short-time dynamics
highlighting the fast oscillating component, see text.} 
\end{figure}
We have numerically solved the driven Bloch-Redfield equation. The real-time dynamics is illustrated in fig.\ \ref{fig:rabiflop}. 

The dynamics shows distinct features on different time scales. As expected,
there are clear Rabi oscillations on the scale of the effective driving
strength (see below). In quantum computing applications, these would
be used for the implementation of a Hadamard gate. 
On top of this, there are fast components:
The dominating one oscillates with the driving frequency,
which originates in the fact that the driving is not perpendicular to 
the static field. 
A weaker one, which oscillates at twice the driving frequency, comes
from the counter-rotating term perpendicular to the static field. 
These oscillations can lead to errors of the Hadamard gate. 
On a longer time scale, the Rabi oscillations decay. The 
time scales will be discussed later on. In general, if one is not exactly
on resonance, these oscillations are combined with nonoscillatory decay, 
see fig.\ \ref{fig:rabiflop} a) and b).
At very long times, 
the system assumes a 
quasistationary value $P_\infty$. 

\begin{figure}[htb]
\includegraphics[width=\columnwidth]{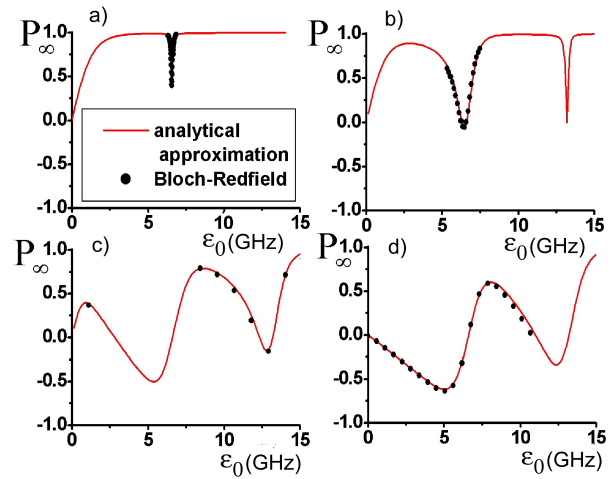}
\caption{\label{fig:spectrotrace}
$\langle\sigma_z\rangle$ in the long-time limit at fixed 
frequency $\Omega/2\pi=6.6$GHz as a function of the energy bias $\epsilon_0$
for different values of the driving strength $s/\Omega=0.034, 0.43, 1.7, 2.4$ 
(a to d). From b to d, nonlinear resonances can be identified. d already
shows negative values at {\em small positive} $\epsilon_0$, which can be 
identified
as the coherent destruction of zero-photon tunneling. Further peaks 
occur at even higher bias.}
\end{figure}

\begin{figure}[htb]
\includegraphics[width=\columnwidth]{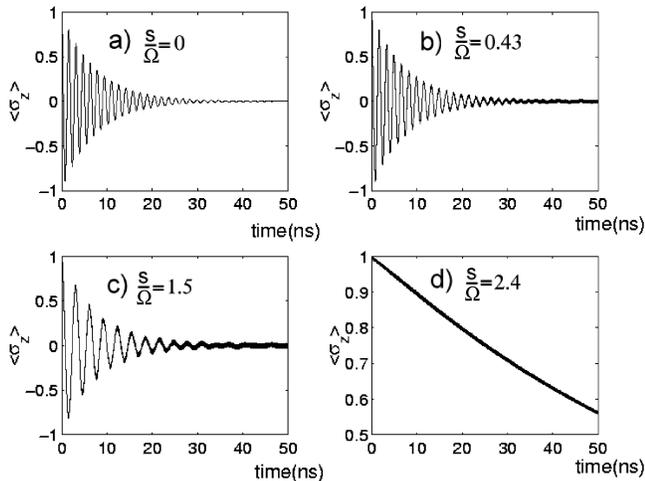}
\caption{\label{fig:stoptunneling}
$\langle\sigma_z\rangle$ at strong driving
with high frequency, $\Omega/2\pi=6.6$ GHz
(where $\Delta/2\pi=660$ MHz)
By increasing
the driving strength, the tunneling is slowed down and brought to a standstill.
}
\end{figure}

Corresponding to the situation of 
a spectroscopy experiment, we now turn to the analysis of
 the quasi-stationary state which
is established after a long time $t\gg \tau_\phi, \tau_R, \omega_R^{-1}$. 
We compare our full numerical solutions with analytical
expressions
we have obtained from the high-frequency approximation
of Refs.\ \onlinecite{LHartmann,Sasetti}. As a result of this approach, the
TSS is mapped onto a coupled ensemble of TSSs corresponding 
to the original system emitting or absorbing $n$ photons from the driving
field during the tunneling. The energy bias of these individual
 systems is $\epsilon_n=\epsilon_0-n\Omega$ and the tunnel matrix element
\begin{equation}
\Delta_n=\Delta J_n(s/\Omega)
\label{eq:deltan}
\end{equation}
where the $J_n$ are Bessel functions. 
At low driving fields, we can approximate 
$\Delta_n=\frac{\Delta}{n!}(s/2\Omega)^n$ as we would expect from the
expansion of a perturbation series in the driving strength. The $\Delta_n$ 
can hence be viewed as
$n$-photon Rabi frequencies.
This implies, that the usual
single-photon frequency gets replaced by $\Delta_1\simeq s \Delta/\nu$,
which can be interpreted as only the projection of the driving field
onto the direction in pseudospin space orthogonal to the static Hamiltonian.
In order to obtain the solid curves in 
fig.\ \ref{fig:spectrotrace} the secular equations for 
the Eigenfrequencies have been solved, taking into account 
an appropriate number of terms \cite{Report}. 
The dynamical two state systems are characterized by individual
dynamical dephasing 
rates $\Gamma_{\phi,n}$ and a common relaxation rate $\Gamma_r$
\cite{LHartmann}. On the n-th resonance,
$\Gamma_{\phi,n}$ can be
very low, much lower than off-resonance,
 as can be seen in figure \ref{fig:rabiflop}, and largely exceed 
the intrinsic dephasing time. This has been observed 
in Refs.\ \onlinecite{Vion,Irinel}.
Figure \ref{fig:spectrotrace} shows numerical and analytical results for
$P_\infty=\lim_{t\rightarrow\infty} \langle\sigma_z\rangle$ at a fixed 
frequency $\Omega/2\pi=6.6$GHz 
as a function of the energy bias $\epsilon_0$. This
corresponds to a realistic experimental situation \cite{Caspar}. In 
fig.\ \ref{fig:spectrotrace} a), taken at weak driving field, 
only the regular resonance corresponding to 
the transition between the two Eigenstates driven by absorbing a single
photon can be seen. At somewhat stronger driving, fig.\ \ref{fig:spectrotrace} b),
this peak grows wider and a second resonance appears, 
corresponding to the simultaneous
absorption of two photons. At higher fields, fig.\ 
\ref{fig:spectrotrace} c), these peaks grow and start to dominate over the 
background. They also turn asymmetric. This trend culminates in the 
situation shown in fig.\ \ref{fig:spectrotrace} d). In that case, 
$P_{\infty}$ does {\em not} grow to positive values at small positive 
$\epsilon_0$, but
 it gets negative and then 
directly approaches the first resonance. The reason 
for this behavior can be identified within the high-frequency approximation:
The lowest order-tunnel frequency $\Delta_0=\Delta J_0(s/\Omega)$ vanishes
at this particular driving strength. Indeed, comparing figs.\ \ref{fig:spectrotrace} a)
and d) one can see, that the step which is at $\epsilon_0=0$ in case a) is
shifted to $\epsilon_0\simeq\Omega$ in case d). This phenomenon,
the coherent desctruction of tunneling \cite{GrifoniHanggi} relies on
destructive interference of the dressed state \cite{Haroche} 
formed by the TSS and
a cloud of photons from the driving field. This interpretation
is supported by the dynamics of $\langle\sigma_z\rangle (t)$. As seen in 
fig.\ \ref{fig:stoptunneling}, which shows the dynamics at the degeneracy point
for different driving strengths, the zero-photon tunneling is slowed down and
brought to a standstill. If that strong
driving can be applied to solid-state qubits, it would provide an alternative
for controlling $\Delta_0$ by a cw microwave field instead of an additional
magnetic flux as proposed in Ref.\ \onlinecite{Hans}.
\begin{figure}[h]
\includegraphics[width=0.8\columnwidth]{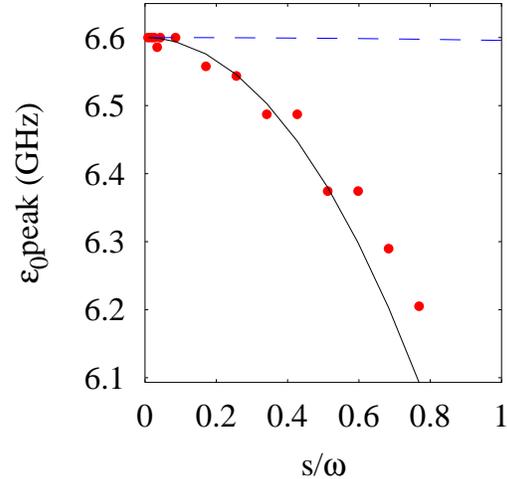}
\caption{\label{fig:peakpos}
Shift of the spectroscopy peak as a function of the driving strength for  
$\omega/2\pi=6.6$GHz and $\Delta/2\pi$=660 MHz. We 
compare to the usual Bloch-Siegert shift formula (dashed) and the formula
derived in the text (solid).}
\end{figure}
At very weak driving, the peak position corresponds to the qubit 
eigenfrequency, $\Omega=\nu$. This is not
reliably predicted by the high-frequency approximation.
At stronger driving, the peak gets shifted. Closer inspection as in 
figure 
\ref{fig:peakpos} shows, that this shift goes much beyond the usual
Bloch-Siegert shift \cite{GrifoniHanggi} of the dynamical 
Eigenfrequency, in fact, 
one can show that the position of the {\em peak} in steady state
and the Eigenfrequency do not conincide. The former is given by balancing
of rates and it can be shown that in lowest order
gets shifted by\cite{Report} $\delta\epsilon_{\rm peak}\simeq s^2/8\Omega^2$ whereas
the Bloch-Siegert shift for our case is
$\delta\epsilon_{\rm BS}\simeq \Delta^2s^2/(16\Omega^3)$. 
As a more general conclusion, already at modest not-too-weak
driving, the resonance 
positions do not necessarily reflect the Eigenfrequencies
of the system. 
\begin{figure}[h]
\includegraphics[width=0.8\columnwidth]{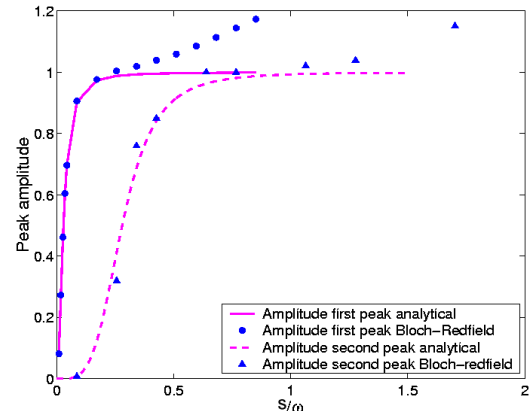}
\caption{\label{fig:amplitude}
Heights of the two lowest resonances as a function of the driving 
strength $s$ at $\nu/2\pi=6.6$ GHz, 
$\Delta/2\pi=660$ MHz,
$\Omega/2\pi=6.6$ GHz
The solid and dashed lines are
extrapolated NMR formulae.}
\end{figure}
In fig.~\ref{fig:amplitude}, the height of the two lowest order peaks 
is shown. It can be seen,
 that, 
from the low-driving side, they 
saturate as soon as their effective Rabi frequency $\Delta_n$ 
exceeds $1/\sqrt{T_rT_\phi}$. 
At 
very high driving, the peaks show an inversion of population.

For the optimization of qubit setups on the way to coherent dynamics, it
is important to characterize its coherence properties from the spectroscopic
data. In NMR, this is done from the linewidth given by
\begin{equation}
\delta\Omega=2\sqrt{\tau_\phi^{-2}+\omega_R^2\tau_R/\tau_\phi}
\label{eq:nmrwidth}
\end{equation}
where $\omega_R$ 
is the Rabi frequency at resonance, which coincides with the
strength of the driving field. 
A generalization of this formula to our case has to 
take into account low temperatures 
and the different
driving situation. Moreover, $\omega_R$ is usually not directly known to
sufficient precision, because the driving strength
 depends on the attenuation of the 
applied fields on their way to the sample and the efficiency of the coupling
\cite{Caspar,CasparFrankEPJ}. 
\begin{figure}[htb]
\includegraphics[width=\columnwidth]{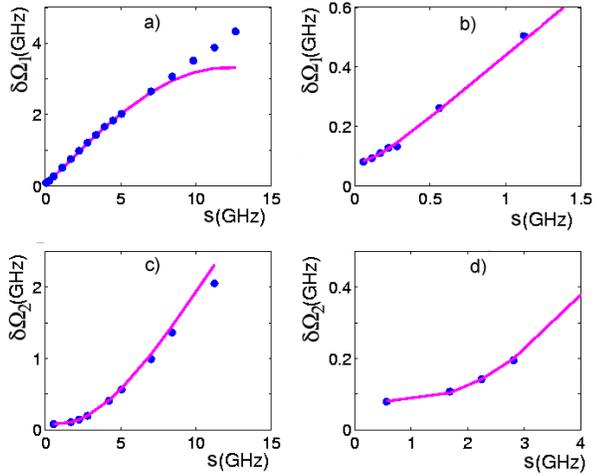}
\caption{\label{fig:width}
Widths of the two lowest resonances as a function of the driving 
strength $s$ at 
 $\Delta/2\pi=660$ MHz
$\Omega/2\pi=6.6$ GHz.
The solid
line corresponds to the extrapolated NMR formula discussed in the text.}
\end{figure}
Our analysis suggests the 
generalization of eq.\ \ref{eq:nmrwidth} is given by
\begin{equation}
\delta\Omega_n=2\sqrt{\tau_\phi^{-2}+\Delta_n^2\tau_R/\tau_\phi}\quad 
n=\pm1,\pm2\dots
\label{eq:fwhm}
\end{equation}
where $\delta\Omega_n$ is the width (in frequency) of the $n$-photon 
resonance and $\Delta_n$ is the effective Rabi frequency defined above. 
At low powers 
$s<\Omega$, they are given by the rates from the undriven ohmic case 
\begin{equation}
\tau_{\rm R}^{-1}=\alpha\frac{\Delta^2}{2\nu}\coth\left(\frac{\hbar\nu}{2k_BT}\right)\quad
\tau_{\rm \phi}^{-1}=(2\tau_R)^{-1}+2\pi\alpha \frac{k_B T}{\hbar}\frac{\epsilon_0^2}{\nu^2}.
\label{eq:dcscales}
\end{equation}
This result is confirmed by our numerical simulations fig. 
\ \ref{fig:width}. 
We can essentially identify three regimes: A saturation broadening regime
at low powers, where $\delta\Omega_n\simeq2\tau_{\rm\phi}^{-1}$, a 
saturated regime, $\delta\Omega_n\simeq2\Delta_n\sqrt{\tau_R/\tau_\phi}$ and a 
nonlinear regime, where the numerical curve deviates from eq.\ \ref{eq:fwhm} 
due to the fact, that the high Rabi frequency shifts the relevant energy
scales and modifies the timescales given in eq.\ \ref{eq:dcscales}. Note, 
that in this regime, the general curve of $P_\infty$ is greatly deformed 
(see fig.\ \ref{fig:spectrotrace}) and
the width of a peak becomes ambiguous. 

This result allows to measure essentially all interesting parameters
of the system experimentally. By extrapolating the level separation
at the degeneracy point (as it was done in \onlinecite{Caspar}), one 
obtains $\Delta$. By tracking the resonance positions at weak driving,
one can evaluate $\epsilon_0$ as a function of the external control parameter
(in \onlinecite{Caspar} this would be the magnetic flux). By driving in 
the saturated regime, the widths of the first and second peak become,
according to eq.\ \ref{eq:fwhm} 
$\delta\Omega_{1/2}=2\Delta_{1/2}\sqrt{\tau_R/\tau_\phi}$, hence by taking their
ratio we find the effective driving strength from 
$\Delta_2/\Delta_1=J_1(s/\Omega)/J_0(s/\Omega)\simeq s/2$ and by tracking
the slope of the first resonance we find the ratio $\tau_r/\tau_\phi$. 
Finally, examining the saturation broadening regime of the first resonance
gives the absolute value of $\tau_\phi$. 

In conclusion, we have numerically and analytically analyzed the
spin-boson system, which e.g.\ represents a SQUID-qubit,
 in the weak
damping regime, driven by continuous fields. As compared to the more
familiar situation in NMR, this system is both different in the character of 
the driving and the low temperature governing the dissipation. We have
shown, that the key features of this system, Rabi-oscillations, and 
saturation of the linewidth, persist qualitatively as has been experimentally
confirmed \cite{Irinel}. They are however altered
on a quantitative level, such as an unanticipatedly strong shift of the
position of the resonance peak,
 and also supplemented by new phenomena such as
higher-harmonics generation, oscillations of $\langle\sigma_z\rangle$ 
on the scale of 
the driving field, and coherent desctruction of tunneling. We have finally
outlined a scheme how to determine all
relevant parameters (tunnel splitting, energy dispersion, driving 
strength, dephasing and relaxation time) 
of a quantum bit solely through spectroscopy. 

We thank M.\ Grifoni, C.H.\ van der Wal, A.C.J.\ ter Haar, C.J.P.M.\ Harmans, 
J.\ von Delft
and
I.\ Goychuk for 
discussions. Work supported by the EU through TMR ``Supnan'' and
IST ``Squbit''. FKW acknowledges support by the ARO under contract-Nr.
P-43385-PH-QC.

\end{document}